\documentclass[pre,preprint,nofootinbib,showpacs]{revtex4}

\usepackage{amsmath}
\usepackage{amssymb}
\usepackage{epsfig}

\def\be{\begin{equation}}
\def\en{\end{equation}}

\newcommand{\EE}[1]{\left<#1\right>}

\def\RR{\rm \hbox{I\kern-.2em\hbox{R}}}
\def\NN{\rm \hbox{I\kern-.2em\hbox{N}}}
\def\ZZ{\rm {{Z}\kern-.28em{Z}}}
\def\CC{\rm \hbox{C\kern -.5em {\raise .32ex \hbox{$\scriptscriptstyle
|$}}\kern
-.22em{\raise .6ex \hbox{$\scriptscriptstyle |$}}\kern .4em}}

\def\<{\langle}
\def\>{\rangle}

\begin{document}

\title{Intermittency of surface layer wind velocity series in the mesoscale range}

\author{Jean-Fran\c{c}ois Muzy}
\email{muzy@univ-corse.fr}
\affiliation{CNRS UMR 6134, Universit\'e de Corse, Quartier Grossetti,
20250, Corte, France}

\author{Rachel Ba\"{\i}le}
\email{baile@univ-corse.fr}
\affiliation{CNRS UMR 6134, Universit\'e de Corse, Vignola,
20200, Ajaccio, France}

\author{Philippe Poggi}
\email{poggi@univ-corse.fr}
\affiliation{CNRS UMR 6134, Universit\'e de Corse, Vignola,
20200, Ajaccio, France}

\date{\today}

\begin{abstract}
We study various time series of surface layer wind velocity at different locations
and provide evidences for the intermittent
nature of the wind fluctuations in mesoscale range.
By means of the magnitude covariance analysis, which is shown to be a more efficient tool to study intermittency than classical scaling analysis, we
find that all wind series exhibit similar features than those observed
for laboratory turbulence. Our findings suggest the existence of a ''universal'' cascade mechanism associated with the energy transfer between synoptic motions and turbulent
microscales in the atmospheric boundary layer.
\end{abstract}


\maketitle

\section{Introduction}
Atmospheric surface layer motions are
a source of many challenging problems.
The issue of designing a faithful statistical model of
spatio-temporal wind speed fluctuations has been addressed
in various fields like boundary layer turbulence phenomenology,
meteorology, wind power control and prediction or
climatology.
From turbulent gusts to hurricanes, breezes to geostrophic wind, the wind
process is characterized by a wide range of spatio-temporal scales
and all the above mentioned disciplines mainly focus on a specific
subrange of scales.
The modeling of wind speed behavior in the mesoscale range is of
great interest for example in wind power generation or in order to control polluant dispersion. In this range of scales extending
from few minutes to few days ($\sim$ 1-1000 km), the properties of wind velocities are less known than
in the range of planetary motions (synoptic scales) or turbulent motions (microscales) \cite{Fri95,dynmeteo}.
From a physical point of view, because of the importance
of boundary conditions, the heterogeneous and non-stationary nature of the processes involved, it is well admitted that mesoscale wind regimes strongly depend upon various factors like atmospheric conditions, the nature of the terrain and may involve periodic variations (caused by
diurnal temperature variations).
Unlike microscale Kolmogorov homogeneous turbulence, mesoscale fluctuations are therefore not expected to possess any degree of universality \cite{lmsa}.
However, during past few years, some papers have been devoted
to the analysis of scaling laws and intermittency features at large scales
in many geophysical signals like temperature, rainfall
or wind speeds  \cite{lss,Gupta1,Gupta2}.
In Ref. \cite{lmsa}, the authors showed that surface layer wind velocities
recorded at low frequency using a cup anemometer display multiscaling
properties very much like in the high frequency turbulent regime.
Moreover, they claimed that random cascade models
could be pertinent to reproduce the observed intermittent fluctuations.
Along the same
line, in \cite{kava04}, a Multifractal Detrended Analysis was performed
on four different hourly wind speed records and revealed some multiscaling properties of the
series. Even if these studies did not go beyond simple scaling exponent estimation and did not consider problems related to the statistical significance of the obtained results, they had the merit to address questions about
the possible intermittent nature of wind variations in the mesoscale range.
As reviewed below, one of the consequences of multiscaling and intermittency is that small scales fluctuations are strongly non Gaussian and characterized by "bursty" behavior. In Refs. \cite{bobape06,KhoMoTsi07},
such features have been precisely observed on wind variations statistics at
largest microscales (or at smallest mesoscales) and have been shown to be
related to a ``fluctuating'' log-normal turbulent intensity.

In this paper we suggest that this ''fluctuating turbulent intensity'' results from a random cascading process initiated at a time scale of few days. Our aim is to show that, in some sense,
''turbulent'' cascade models are likely to be pertinent at larger scales,
in the so-called mesoscale regime.
As compared to the previously cited papers,
our analysis relies upon the use of magnitude (i.e. logarithms of velocity increments amplitudes)
correlation functions. From
a mathematical point of view, long-range correlated magnitudes have been
shown to be at the heart of the construction of continuous cascade processes \cite{ArnMuzSor98,MuDeBa00}.
For a practical point of view, magnitude covariance possesses interesting properties as far as statistical estimation problems are concerned \cite{DelMuArn01,BacKozMuz06,Koz06}.
We show that these correlation functions can be reliably estimated and
are very similar to those associated with longitudinal velocity time series of laboratory turbulent experiments.
Moreover we observe some universal features among the various analyzed series.

The paper is organized as follows:
in section \ref{s_ma}, we make a brief review of intermittency and the related notion
of random cascade models. We emphasize on the interest of studying magnitude correlations
and discuss its relationship with multiscaling properties. We then compare,
on an empirical ground, the relative performances of intermittency estimators relying upon
scaling and magnitude covariance. This section ends with a review of a recent work
of B. Castaing that has shown how intermittency in Lagrangian and Eulerian frames
is "observed" on time series recorded at a fixed spatial position.
Our main experimental results are then presented in section \ref{s_sa}. After a rapid description of
various wind data we have studied, we show that wind surface layer variations in mesoscale range have intermittent properties
and possess "universal" magnitude covariance similar to laboratory turbulent fluctuations.
Discussions and prospects are reported in section \ref{s_cp}.

\section{Intermittency: from multiscaling to magnitude correlations}
\label{s_ma}

\subsection{Multiscaling and intermittency}
\label{s_ma1}

Small scale intermittency is one of the most challenging problems in contemporary turbulence research.
It is generally associated with two distinctive features: the first one is that at small scales the probability density functions (pdf) of velocity variations are strongly leptokurtic (with ''stretched exponential" tails)
while they are almost gaussian at larger scales.
The second one is that the so-called structure functions display multiscaling
properties. As we shall see, these two properties are in fact
equivalent within the multiplicative cascade picture.
Let us make a brief overview of these notions.

We denote $X(t)$ a continuous process (for instance the time variations
of a velocity field component at a fixed location)
and let $\delta_\tau X(t)$ be its increments over a scale $\tau$: $\delta_\tau X(t) = X(t+\tau)-X(t)$.
One usually defines the order $q$ structure function of $X$ as
\begin{equation}
\label{defsq}
          S_q(\tau) = \int |\delta_\tau X(u)|^q \; du
\end{equation}
and the $\zeta(q)$ spectrum as the scaling exponent of $S_q(\tau)$:
\begin{equation}
\label{defzeta}
    S_q(\tau) \operatornamewithlimits{\sim}_{\tau \rightarrow 0} \tau^{\zeta(q)}
\end{equation}
If the function $\zeta(q)$ is non-linear one says that $X(t)$
is a multifractal process or an intermittent process.
In that case, as shown e.g. in \cite{DelourPHD}, $\zeta(q)$ is
necessarily a concave function and
the previous scaling holds in the range
of small $\tau$ ; $\tau \rightarrow 0$ means precisely
$\tau \ll T$ where $T$ is a coarse scale called the integral scale
in turbulence (in general associated with the injection scale).
The intermittency coefficient is a positive number that quantifies the
non-linearity of $\zeta(q)$ and can be defined \footnote{Notice that one
can find different definitions of the ``intermittency coefficient'' or the ``intermittency exponent'' in the literature (see \cite{AnAnDa01} for example).
However within the log-normal cascade model discussed below they are all equivalent.}  as, e.g.,
\begin{equation}
\label{deflambda}
   \lambda^2 = -\zeta''(0)
\end{equation}
The most common example of non linear $\zeta(q)$ function is the so-called
{\em log-normal} spectrum which is a simple parabola:
\begin{equation}
\label{ln}
  \zeta(q) = \alpha q -\frac{\lambda^2}{2} q^2
\end{equation}
In that case $\lambda^2$ corresponds to the constant curvature of $\zeta(q)$.

In order to estimate the multiscaling properties and/or the intermittency
coefficient, one can directly estimate partition functions from
the data and obtain the exponent $\zeta(q)$ from a least square fit of $S_q(\tau)$ in log-log representation.
However this method suffers from various drawbacks. First, from
a fundamental point of view, one has to distinguish temporal (or spatial) averages from ensemble averages.
For instance, in the case of a log-normal multifractal,
rigorously speaking, only the ``ensemble'' average $\EE{|\delta_\tau X(u)|^q}$ behaves as a power
law with an exponent $\zeta(q)$ as given by Eq. \eqref{ln}. The temporal or spatial average has an exponent spectrum $\zeta(q)$
that becomes linear above some value of $q$ and is no longer parabolic \cite{LahAbrCha04}.
In order to estimate $\zeta(q)$
from moment scaling over a wide range of $q$ one has to use the so-called ``mixed'' asymptotic framework \cite{MuBaBaPo08}.
But the main problem remains that high order moment estimates require very large sample size. Moreover the scaling behavior
can also be altered by finite size effects, discreteness and
non stationarity effects like periodic perturbations or periodic
modulations of the data (see below).
A more reliable method first introduced in \cite{DelMuArn01} (see also \cite{Basu07}) relies upon the so-called
magnitude cumulant analysis. It simply consists in focusing on the scaling
behavior of partition functions around $q = 0$.
According to this approach, the structure function is written as:
\begin{equation}
   S_q(\tau) = \exp \left( \sum_{k=1}^{\infty} C_k(\tau) \frac{q^k}{k!} \right)
\end{equation}

where $C_k(\tau)$ is the k-th cumulant associated with the random variable
$\omega_\tau(u) = \ln(|\delta_\tau X (u)|)$. The logarithm of the increment will henceforth be referred to as the {\em magnitude} of velocity increments.
Note that $C_1$ is simply the mean value of $\omega$ while $C_2$
is its variance. Thanks to the scaling relationship \eqref{defzeta}, one
deduces that all cumulants behave as:
\begin{equation}
  C_k(\tau) = c_k \ln(\tau) + r_k
\end{equation}
where the constants $\{ r_k \}$ account for both the integral scale and the prefactors
in the scaling relationship \eqref{defzeta}.
The function $\zeta(q)$ can therefore be
expressed in terms of a cumulant expansion:
\begin{equation}
\label{cumexp}
  \zeta(q) = \sum_k c_k \frac{q^k}{k!}
\end{equation}
In particular one sees that the intermittency coefficient is directly
involved in the behavior of the magnitude variance as:
\begin{equation}
\label{cumulant2}
   C_2(\tau) = - \lambda^2 \ln(\tau) + r_2 \; .
\end{equation}
Eq. \eqref{cumulant2} has been successfully used to estimate the intermittency
coefficient of longitudinal velocity fields in 3D fully developed regime
under various experimental conditions. A common value close to
$\lambda^2 \simeq 0.025$ has been obtained. It is remarkable
that this intermittency value appears to be
universal \cite{ArnManMuzy98,DelMuArn01,Basu07}.

Let us end this brief review by discussing
the relationship between intermittency and
the small scale ``bursty'' behavior of velocity increments.
Indeed, it is well known in turbulence laboratory experiments
that large scale increment pdf are close to be normal while small scales
have larger tails. The flateness strongly increases when one goes from
large to small scales.
We focus only on the log-normal case, i.e., a parabolic $\zeta(q)$,
extension of our considerations to other laws being straightforward.
Thanks to the structure function multiscaling interpreted as a moment
equality, by simply performing a time scale contraction, $\tau' = s \tau$ ($s<1$),
one can write:
\begin{equation}
\label{ss1}
    \delta_{s \tau} X \operatornamewithlimits{=}_{Law} e^{\Omega_s} \delta_{\tau} X
\end{equation}

where $\Omega_s$ is a Gaussian random variable of variance $-\lambda^2 \ln(s)$
which law is denoted as $G_s(\Omega)$.
By simply assuming that for $\tau = T$ ($T$ being the integral scale),
the increments of $X$ are normally distributed, one obtains the small scale pdf of $\delta_\tau X$ by the
well known Castaing formula \cite{CasGagHop90}:
\begin{equation}
\label{castaing}
   p(z,\tau)  = (2\pi)^{-\frac{1}{2}} \int G_{\tau/T}(\Omega) e^{\Omega} e^{-\frac{e^{-2\Omega}z^2}{2}} \; d\Omega
\end{equation}

This means that, at scale $\tau$, the pdf of the increments of the process $X(t)$
are obtained as a superposition of Gaussian distributions
of variance $e^{2\Omega}$ where $\Omega$ is itself a gaussian random variable of variance increasing at
small scales like $\lambda^2 \ln(T/\tau)$.
The smaller the scale, the larger
the variance of $\Omega$ and therefore the larger the tails of the increment
pdf $p(z,\tau)$. The continuous deformation from Gaussian at large scales
towards stretched exponential like shapes at small scales, observed in laboratory
turbulence experiments, has been shown to be well accounted by the transformation \eqref{castaing}.

\subsection{Random cascades and logarithmic magnitude covariance}
\label{sec_casc}
A natural question that arises after the previous analysis is how
can we explicitely build models that possess multiscaling properties?
In other words, since multiscaling is
equivalent to intermittency, how can the variance
of the magnitude increase as a logarithm of the scale as described
by Eq. \eqref{cumulant2} ?
The answer comes from the self-similarity Eq. \eqref{ss1} that can be iterated and interpreted
as a random cascade: when one goes from large to small scales, one multiplies the process by a random variable $W_s = \exp(\Omega_s)$.

Usually one starts by building a non-decreasing (i.e. with positive variations)
cascade process, denoted hereafter as $M(t)$, which is often referred to as a multifractal {\em measure}
though its variations are not bounded.
More general multifractal processes (or multifractal "walks")
can be simply obtained by considering a simple Brownian motion $B(t)$
(or any self-similar random process) compounded with
the measure $M(t)$ considered
as a stochastic time: $X(t) = B\left[M(t)\right]$.
In finance literature $M(t)$ is often referred to as the 'trading time' while
in turbulence $M(t)$ can be associated with the spatial or temporal distribution
of energy dissipation.
The statistical properties of $X(t)$ can be directly
deduced from those of $M(t)$ (see e.g., \cite{MuBa02,BaMu03}). Random multiplicative cascades measures were originally
introduced as models of the energy
cascade in fully developed turbulence. After the early works of
Mandelbrot \cite{Man74a,Man74b,Man03}, a lot of mathematical studies have been devoted to discrete
random cascades \cite{KaPe76,Gui87,Mol96,Mol97,Liu02}. Let us summarize the main properties of these constructions,
set some notations and see how these notions extend to continuous cascades.

The simplest multifractal discrete cascades are the dyadic
cascades defined by the following iterative rule: one starts
with some interval of constant density and splits
this interval in two equal parts. The
density of the two sub-intervals is obtained by multiplying the original
density by two (positive) independent random factors $W$ of same law.
This operation is then repeated {\em ad infinitum}.
The integral scale corresponds to the size of the starting interval.
A log-normal cascade corresponds to $W = \exp(\Omega)$ where $\Omega$ is
normally distributed. Peyri\`ere and Kahane \cite{KaPe76} proved that
this construction converges almost surely towards
a stochastic non decreasing process $M_{\infty}$ provided $\EE{W \ln W} < 1$.
The multifractality of $M_{\infty}$ (hereafter simply denoted as $M$)
and of $X(t) = B(M(t))$  ($B(t)$ being a Brownian motion)
has been studied by many authors (see e.g. \cite{Mol96}) and
it is straightforward to show that the spectrum of $X(t)$ is
\begin{equation}
\label{zetaM}
  \zeta(q) = q/2-\ln_2\EE{W^{q/2}}
\end{equation}
The self-similarity Eq. \eqref{ss1} for dilation factors $s = 2^{_n}$
can also be directly deduced from the construction.

Because the discrete cascade construction involves dyadic intervals
and a 'top-bottom' construction it is far from being stationary.
In order to get rid of theses drawbacks,
continuous cascade constructions
have been recently proposed and
studied on a mathematical ground \cite{MuDeBa00,SchMar01,BaMan02,MuBa02,BaMu03,ChaRieAbr05}.
They share exact multifractal scaling with discrete cascades
but they display continuous scaling and possess stationary
increments \cite{MuDeBa00,BaMan02,MuBa02,BaMu03}.
Without entering into details, we just want to
stress that these constructions rely upon
a family of infinitely divisible random processes $\Omega_l(t)$.
The process $M(t)$ is obtained as the weak limit, when $l \rightarrow 0$, of
$\int_0^t e^{\Omega_l(v)} dv$. In the log-normal case, $\Omega_l$ is simply
a Gaussian process defined by a covariance function that mimics the behavior of discrete cascade.
Indeed, as shown in \cite{ArnMuzSor98,ArnBaMaMu98}, if $X(t) = B(M(t))$ where $M(t)$ is a
discrete cascade as defined previously with intermittency coefficient $\lambda^2$,
then correlation function $\rho(\Delta t)$ of the
magnitude $\omega_\tau(u) = \ln(|\delta_\tau X(u)|)$ decreases slowly as
a logarithm function:
\begin{equation}
\label{magcorr}
\rho(\Delta t) \operatornamewithlimits{=}^{def} Cov \left[\omega_\tau(u), \omega_\tau(u+\Delta t) \right] \simeq -\lambda^2 \ln(\frac{\Delta t}{T+\tau}) ~~{\mbox{for lags}}~ \Delta t \leq T.
\end{equation}
The integral scale $T$ where cascading process ``starts'' can therefore be interpreted
as a {\em correlation length} for the variation log-amplitudes of $X(t)$.
The so-called Multifractal Random Walk (MRW) process introduced in Ref.
\cite{MuzDelBac00} consists
in constructing a log-normal multifractal process $X(t) = B(M(t))$
where $B(t)$ is a Brownian motion and
\begin{equation}
M(t) = \lim_{l \rightarrow 0^+} \int_0^t e^{\Omega_l(u)} \; du
\end{equation}
where the $\Omega_l(u)$ is a gaussian process with a logarithmic covariance as
given in Eq. \eqref{magcorr}. In Fig. \ref{fig_est}(a) and \ref{fig_est}(b) are plotted respectively a sample of a log-normal measure $M(dt)$ ($dt = 1$) and a path of a MRW process $X(t)$ corresponding to $\lambda^2 = 0.025$ and $T = 512$.

The studies devoted to continuous versions of discrete cascades have mainly shown that multifractal processes are related to exponential of
logarithmic correlated random processes. The magnitude covariance function has been proven to be at the heart of the notion of ``continuous cascade''.
As we shall see below, it also allows one to estimate the
intermittency coefficient in a more reliable
way than methods based on scaling properties.

\subsection{Intermittency coefficient estimation issues}
\label{s_ie}

As far as the problem of the estimation of the intermittency coefficient is concerned, it results from previous discussion that this coefficient, originally defined as the curvature of $\zeta(q)$ (Eq. \eqref{deflambda}), can be estimated either from the behavior of magnitude variance across scales (Eq. \eqref{cumulant2}) or from the slope of the time magnitude covariance in lin-log coordinates (Eq. \eqref{magcorr}).

\begin{figure}
\begin{center}
\includegraphics[width=10cm]{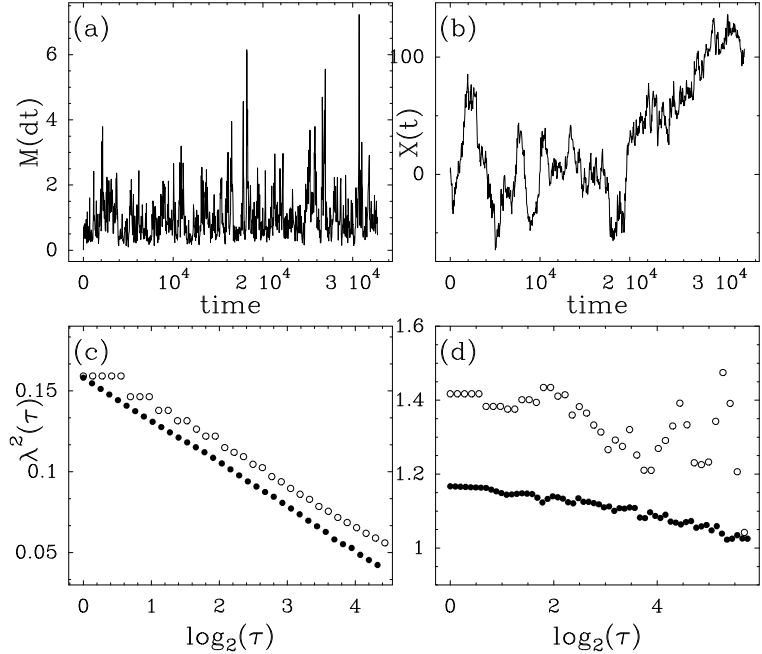}
\end{center}
\caption{
Estimation of the intermittency coefficient for a log-normal
MRW multifractal measure (a) and a log-normal MRW process (b).
In both cases the sample size is 32768 points, $\lambda^2 = 0.025$ and
$T = 512$. In (c) are plotted the magnitude covariance ($\bullet$)
and magnitude variance ($\circ$) as a function respectively of the logarithm of the lag and the logarithm of the scale. The slope of both curves provides an
estimation of $\lambda^2$. One can see that for measure $M(dt)$, the errors
in the estimation are comparable. In (d) are displayed the same plots as in (c) but for
the magnitude of the MRW process. Because of the additive noise and the small
number of independent points at large scales, the estimation relying upon the magnitude variance turns out to be much more altered.
}
\label{fig_est}
\end{figure}

Let us show that this latter definition is much more reliable
than former one. The precise computation of the properties of estimators relying upon Eqs \eqref{cumulant2} and
\eqref{magcorr} is a difficult task. In Ref. \cite{BacKozMuz09},
it has been shown that a Generalized Method of Moments relying on
the magnitude correlation function provides an unbiased and consistent estimator of $\lambda^2$.
Estimators relying on Eq. \eqref{cumulant2} have been precisely discussed within the context of atmospheric turbulence in Ref. \cite{Basu07}
where the authors showed that it allowed one, by means of a bootstrap method called ``surrogate analysis'',
to estimate $\lambda^2$ and distinguish intermittent
from non intermittent time series.

A simple argument allows us to understand why magnitude covariance based estimation (Eq. \eqref{magcorr})
is better than scale magnitude cumulant analysis (Eq. \eqref{cumulant2}) in the case of multifractal random walks. This is illustrated in Fig. \ref{fig_est} where both estimators are compared for a sample of a multifractal random measure
constructed from a log-normal continuous cascade (Fig. \ref{fig_est}(a)) and a MRW process (Fig. \ref{fig_est}(b)). One sees in Fig. \ref{fig_est}(c),
that in the case one handles directly the random cascade measure,
both estimators are roughly equivalent. However, in the case of a MRW process,
the measure is not directly observed since it represents
the stochastic variance of the random walk. In this situation,
the estimates based on magnitude covariance are better than those
obtained from the magnitude variance scaling (Fig. \ref{fig_est}(d)).
This feature can be explained as follows:
if one considers that for continuous cascades processes,
Eq. \eqref{ss1} is an equality in law
for all finite dimensional distributions (f.d.d.), by taking the logarithm, one gets:
\begin{equation}
\label{ss2}
 \omega_\tau(u)  = \ln(|\delta_\tau X (u)|) \operatornamewithlimits{=}_{f.d.d.}   \Omega_\tau(u) + \ln(|\epsilon(u)|)
\end{equation}
where $\Omega_\tau(u)$ is a log-correlated gaussian random variable (Eq. \eqref{magcorr})
which variance behaves like in Eq. \eqref{cumulant2} while $\epsilon(u)$ is a standardized normal white noise independent of $\Omega_\tau$.
Note that in the case of a multifractal measure, there is no noise term $\ln(|\epsilon|)$) and the estimation error is small as compared
to the situation of MRW when this term is present. One can then assume that the estimation error mainly comes from the presence of $\ln(|\epsilon|)$.
This error is thus expected to decrease, in both cases, as $\frac{1}{\sqrt{N_e}}$ if $N_e$ is the number of effective independent samples of the considered signal. This effective number is proportional to the overall sample size but decreases with the scale $\tau$.
Therefore, since at scale $\tau$, the effective number of independent points is $N_e \simeq N \tau^{-1}$, if one
fits Eq. \eqref{cumulant2} over the range of scale extending from $\tau = 1$ to $\tau = \tau_{\max}$,
the mean error is roughly proportional to $\sqrt{\tau_{\max} N^{-1}}$. On the other hand, by fitting Eq. \eqref{magcorr} at scale $\tau=1$ from lags $\Delta t = 1$ to $\Delta t = \tau_{\max}$, one expects
an approximately constant error of $\sqrt{N^{-1}}$.
Moreover, the error in the computation of the variance is
proportional to the square root fourth moment of $\ln(\epsilon)$ while the error in the covariance is proportional to the
second moment. Since the ratio between these two quantities is 3,
the ratio of errors between the two methods is of order:
\begin{equation}
    r \simeq 3 \sqrt{\tau_{\max}}
\end{equation}

Let us notice that, rigorously, the error also depends on the overall range of values used to apply the linear regression and thus one expects,
in the case of an estimation relying upon Eq. \eqref{cumulant2},
there is an optimal range of scale to perform the fit. This optimum results, on the one hand, from a reduction of the error
when one increases the fitting range of scales and, on the other hand,
from the increasing of the error associated with the small number
of independent points at large scales (see Ref. \cite{audit} where such a problem is considered in details).
We have performed extensive Monte-Carlo estimations using synthetic log-normal continuous cascades
and we have checked that for parameters, sample sizes and range of scales close to the experimental values considered below (see section \ref{s_sa}),
the observed ratio between the errors is $r_{exp} \simeq 10$ in agreement
with previous considerations. The intermittency coefficient of a random cascade is thus estimated
in a much more reliable way using magnitude correlation functions.

\subsection{Squared log magnitude covariance as the result of Lagrangian and Eulerian intermittency}
\label{sq_mag}

In most fluid mechanics experiments, one usually records one or several components of the velocity
field using an anemometer, so one has only access to the values of the field at a fixed spatial position as a function of time.
In order to make inferences on the spatial properties of the velocity, i.e., the Eulerian field, one generally invokes the Taylor
Frozen hypothesis or, when the turbulence rate is large, the Tenekes sweeping argument according to which small scale velocity fluctuations are mainly
caused by Eulerian variations advected by large scale random motions \cite{Che05}.
However, as explained below, for some functions, single point measurements cannot be linked so easily to their Eulerian counterpart.
Hereafter, we reproduce the argument of B. Castaing \cite{Cast02,Cast06}
in order to understand the shape of the single point
magnitude covariance if one supposes that both Eulerian and Lagrangian
velocity fluctuations are given by a continuous cascade as described previously.

Let us denote $\Omega(x,t)$ the magnitude at time $t$ and position $x$. It is important to notice that $\Omega(x,t)$ is a local
field that does not depend on any spatial or time scale (for example in turbulence, $\Omega(x,t)$ can be considered as the logarithm of the
dissipation field or of the velocity increments at a scale smaller than the Kolmogorov dissipation scale).

Our goal is to compute
\begin{equation}
     \mbox{Cov} \left[\Omega(x,t),\Omega(x,t+\Delta t)\right]
\end{equation}
as a function of the time lag $\Delta t$ for a fixed value of $x$.

If one supposes that both Eulerian and Lagrangian fields $\Omega$ are
log-correlated, i.e., well described by a continuous cascade model, then
\begin{eqnarray*}
     \mbox{Cov}\left[\Omega(x,t),\Omega(x+r,t)\right] & = & \lambda^2 \ln(\frac{L}{r+\eta_e}) \\
     \mbox{Cov} \left[\Omega(x(t),t),\Omega(x(t+\Delta t),t+\Delta t)\right] & = &  \mu^2 \ln(\frac{T}{\Delta t+\eta_l}) \\
\end{eqnarray*}
where $\Delta t$ and $r$ are time and space lags, $\eta_e$ and $\eta_l$ are small scale spatial and temporal cut-offs (i.e. the
Kolmogorov scales in turbulence), $L$ and $T$ are spatial and temporal integral scales
and  $\lambda^2$, $\mu^2$ are Eulerian and Lagrangian intermittency coefficients.

Let us suppose that the fluid particle at position $x$ at time $t+\Delta t$ was
at position $x'$ at time $t$.
Thanks to the above covariance formula, one can write:
\begin{equation}
     \Omega(x,t+\Delta t) = \rho_1 \Omega(x',t)+\epsilon
\end{equation}
where $\epsilon$ is a random variable independent of $\Omega(x',t)$ and:
\begin{equation}
   \rho_1 = \frac{\mu^2 \ln(\frac{T}{\Delta t+\eta_l})}{\mbox{Var} (\Omega)}
\end{equation}

But (at least from a statistical point of view) $r = |x'-x| = V \Delta t$
where $V$ is a 'typical' velocity (the mean or r.m.s. velocity) so that, if one assumes
that $L \simeq V T$:
\begin{equation}
     \Omega(x',t) = \rho_2 \Omega(x,t)+\epsilon'
\end{equation}
where $\epsilon'$ is a random variable independent of $\Omega(x,t)$ and:

\begin{equation}
   \rho_2 = \frac{\lambda^2 \ln(\frac{VT}{V\Delta t + \eta_e})}{\mbox{Var}(\Omega)}
\end{equation}

Then, if $\epsilon$ and $\epsilon'$ are uncorrelated, since the correlation coefficient between $\Omega(x,t)$ and $\Omega(x,t+\Delta t)$ is the product
$\rho_1 \rho_2$, we have

\begin{equation}
     \mbox{Cov} \left[\Omega(x,t),\Omega(x,t+\Delta t)\right] = \mbox{Var}(\Omega) \rho_1 \rho_2 \;,
\end{equation}

and by taking into account the fact that $Var(\Omega_s) = \mu^2 \ln(T/\eta_l) = \lambda^2 \ln(L/\eta_e)$
one obtains finally, for lags such that $V \Delta t \gg \eta_e$ and $\Delta t \gg \eta_l$:

\begin{equation}
    \mbox{Cov} \left[\Omega(x,t),\Omega(x,t+\Delta t)\right] \simeq \frac{\lambda^2}{\ln(\frac{T}{\eta_l})} \ln^2\left(\frac{\Delta t}{T}\right) = \frac{\mu^2}{\ln(\frac{L}{\eta_e})} \ln^2\left(\frac{\Delta t}{T}\right)\; .
\end{equation}

If one plots $\sqrt{\mbox{Cov}}$ as a function of $\ln(\tau)$ one expects
a straight line of slope:
\begin{equation}
     r = \sqrt{\frac{\lambda^2}{\ln \frac{T}{\eta_l}}} = \sqrt{\frac{\mu^2}{\ln \frac{L}{\eta_e}}}
\end{equation}

In one knows the ``Reynolds numbers'' $\frac{L}{\eta_e}$ and $\frac{T}{\eta_l}$, the intermittency coefficients
can be estimated from the slope $r$ as:
\begin{eqnarray}
\label{l2sq}
          \lambda^2 & = & r^2 \ln(\frac{T}{\eta_l}) \\
          \mu^2  & = & r^2 \ln(\frac{L}{\eta_e})
\end{eqnarray}

We see that, by taking into account both Eulerian and Lagrangian fluctuations
on single point measurements, one should observe a squared logarithm magnitude covariance instead of the logarithmic behavior of Eq. \eqref{magcorr}.
This peculiar shape of magnitude covariance has indeed been precisely observed on laboratory fully developed turbulence data
in Ref. \cite{DelMuArn01}.

\section {Results: intermittency in mesoscale wind fluctuations}
\label{s_sa}

\subsection{The data series}
The results reported in the following are based on different wind
velocity time series. The first data set consists in horizontal wind
speeds and directions that have been recorded
every minutes during 5 years (1998-2002) at our laboratory in
Ajaccio - Vignola at a height of $10$ m
by means of a cup anemometer.
We also study hourly wind speed and direction data (10 minutes averages)
for 7 different sites in Corsica (France). The length
of these series is also 5 years.
These data have been measured and collected
by the French Meteorological Service of Climatology (Meteo-France)
using a cup anemometer and wind vane at $10$ meters above ground level.
Finally, we also consider potential winds from
KNMI HYDRA PROJECT available online \cite{KNMI}: they represent series of hourly (1 hour average), $
10$ meters potential wind speed gathered
during several years at various locations in Netherlands.
More specifically we consider the series recorded at 3 different sites over
a long period of several decades.
Table \ref{table1} summarizes the main characteristics of the studied series.
\begingroup
\squeezetable
 \begin{table}[h]
\begin{center}
 \begin{tabular}{|c|c|c|c|c|c|}
  \hline
   Location & Latitude & Longitude & Dates & Sampling freq. & Site\\
  \hline
   Vignola (Ajaccio) & $41^o56$'N & $8^o54$'E & 1998-2003 & 1 min & 50m, coastal, high hills\\
  \hline
   Ajaccio & $41^o55$'N & $8^o47$'E & 2002-2006 & 1 hour & 5m, coastal, plain, airport \\
  \hline
   Bastia & $42^o33$'N & $9^o29$'E & 2002-2006 & 1 hour & 10m, coastal, plain, airport \\
  \hline
   Calvi & $42^o31$'N & $8^o47$'E & 2002-2006 & 1 hour & 57m, coastal, hills\\
  \hline
   Conca & $41^o44$'N & $9^o20$'E & 2002-2006 & 1 hour & 225m, high hills\\
  \hline
   Figari & $41^o30$'N & $9^o06$'E & 2002-2006 & 1 hour & 21m, plain, airport, hills\\
  \hline
   Renno & $42^o11$'N & $8^o48$'E & 2002-2006 & 1 hour & 755m, mountains\\
  \hline
   Sampolo & $41^o56$'N & $9^o07$'E & 2002-2006 & 1 hour & 850m, mountains\\
  \hline
   Eindhoven & $51^o44$'N & $5^o41$'E & 1960-1999 & 1 hour & 20m, plain\\
  \hline
   Ijmuiden & $52^o46$'N & $4^o55$'E & 1956-2001 & 1 hour & 4m, coastal, plain\\
  \hline
   Schipol & $52^o33$'N & $4^o74$'E & 1951-2001 & 1 hour & -4m, plain, airport \\
  \hline
\end{tabular}
\end{center}
\caption{Main features of the time series}
\label{table1}
\end{table}
\endgroup

In the sequel, $v(t)$ will denote the modulus of the velocity horizontal vector
while $v_x(t)$ and $v_y(t)$ will stand for its two components along arbitrary
orthogonal axes $x$ and $y$. We have by definition:
\[
   v(t) = \sqrt{v_x(t)^2+v_y(t)^2}
\]
Because there is no well defined constant mean velocity direction with a small turbulent rate, we
have chosen to study $v_x$ and $v_y$ separately instead of considering meaningless longitudinal and transverse
components.

\begin{figure}
\begin{center}
\includegraphics[width=10cm]{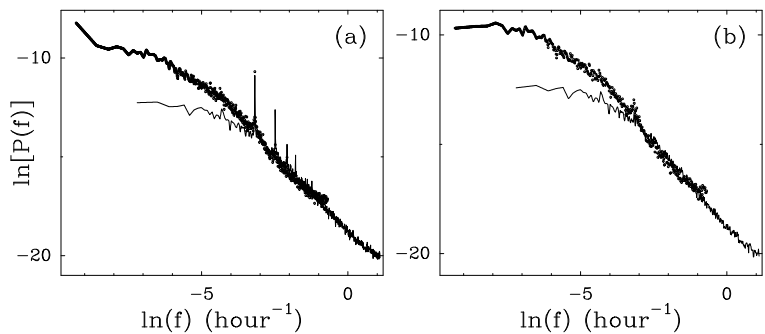}
\end{center}
\caption{
\label{figSpectrum}
Power spectrum density of $v_x$ wind components 10 min rate Vignola series (solid line) and hourly Eindhoven series (dotted line)
in log-log representation. (a) One clearly identifies the peaks associated
with diurnal oscillation superimposed to an overall scaling regime where
$P(f) \sim f^{-\beta}$ with $\beta \simeq 1.6$.
This regime roughly extends from few days to few minutes.
Turbulent atmospheric scales are below
the finest resolved time scale. (b) Plot of the same spectra where the estimated local seasonal components have been removed.}
\end{figure}

The power spectrum analysis is one of the most common tools
for analyzing random functions and is at the heart of a wide number
of studies of wind velocity statistics.
Since the pioneering work of Ven Der Hoven \cite{vdh,ot},
the shape of a typical atmospheric wind speed spectrum in the atmospheric
boundary layer is still matter of debate. It is relatively well admitted
that it possesses two regimes separated by low energy valley called the
``spectral gap'' located at frequencies around few
minutes. This gap separates the microscale regime, where turbulent
motions take place, from the mesoscale range. In the homogeneous turbulent
regime, it is well known that the spectrum associated with the
velocity field behaves like $E(k) \sim k^{-5/3}$ as predicted
by Kolmogorov in 1941 \cite{Fri95} ($k$ is the spatial wave-number).
In the mesoscale range, for
time scales greater than few minutes, the
shape of this spectrum appears to depend on various factors.
If some experiments suggest that a $k^{-5/3}$ spectrum can
extend up to synoptic scales \cite{NasGage83,Lilly83} in the free atmosphere,
things are different in the surface layer \cite{kaimal}. Some authors
suggest that statistics a priori depend on local conditions (orographic,
atmospheric,...) and one does not expect the same
degree of universality as in the microscale regime. For example, in
ref \cite{lmsa}, it is shown that the spectrum exponent may depend
on the atmospheric stability conditions and also on the topography.
\begin{figure}
\begin{center}
\includegraphics[width=8cm]{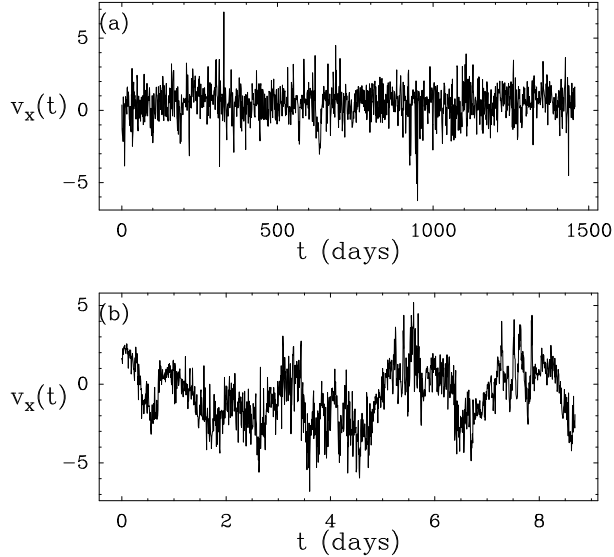}
\end{center}
\caption{Time fluctuations of the $v_x(t)$ component of the Vignola velocity
field. (a) At large time scales, one does not see any structure, we
are in the flat (white noise) regime of the power spectrum. (b)
At a finer time resolution, the series appears as a superposition of
turbulent gusts and a more regular component.}
\label{figvx}
\end{figure}

In Fig. \ref{figSpectrum} are plotted the power spectrum of the $v_x$ component of two series: the "high frequency" (10 min rate) series
of Vignola and the hourly series recorded at Eindhoven.
These two spectra that do not cover the same frequency range have been shifted
by an arbitrary multiplicative factor.
Series associated with the $v_y$ component or corresponding to other sites have similar features.
One can see that, up to the mains peaks associated with diurnal
wind oscillations (see below), these spectra are well described by a power-law $P(f) \sim f^{-\beta}$ in a frequency domain
which corresponds to time scales from few minutes to a characteristic
time of few days (note that the low frequency behavior is much more reliable in the Eindhoven series since it covers a period close
to ten time longer than the Vignola series). The value of the exponent is $\beta \simeq 1.6$ for both sites.
Within the framework of self-similar Gaussian processes \cite{Taqqu},
the fact that $\beta > 1$ for all series indicates that the signal
process $v_x(t)$ has continuous pathes. Since at low frequency the spectrum becomes flat, this means
that the process appears regular at small scales and more ``noisy" at larger scales.
This feature is illustrated in Fig. \ref{figvx} where we have plotted two samples of the component $v_x(t)$
of Vignola wind series over two different time intervals.
At very large scale, the signal looks like a highly irregular
''white noise'' that corresponds to the power-spectrum
low frequency flat behavior  (Fig. \ref{figvx}(a))
while a zoom over a finer time interval reveals more regular variations (Fig. \ref{figvx}(b)).
Notice that one can also observe a daily oscillating behavior and high frequency turbulent gusts superimposed
to these regular random variations.
\begin{figure}
\begin{center}
\includegraphics[width=12cm]{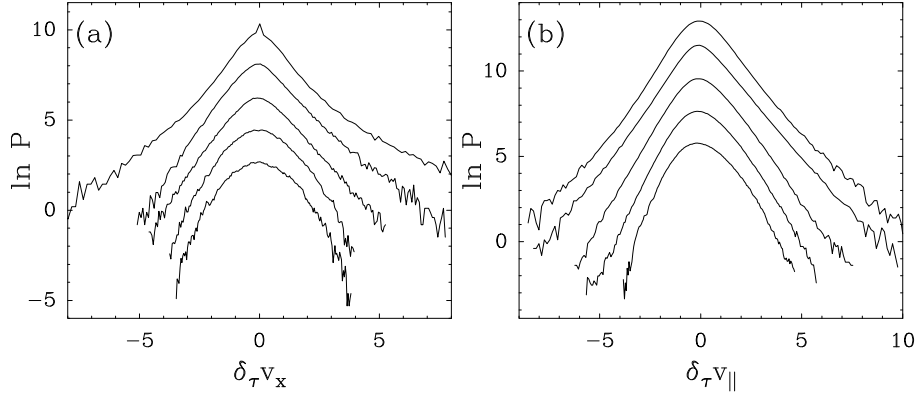}
\end{center}
\caption{
\label{figpdf} Semi-logarithmic representation of standardized velocity increment pdf at various scales. From top to bottom one goes
from small to coarse time scales. All graphs have been vertically shifted for sake of clarity. (a) Logarithm of the pdf of the wind velocity increments $\delta_\tau v_x$ for the Eindhoven wind series.
Time scales $\tau$ go from 1 hour to 5 days.
(b) Same plots as in (a) but for the increments of the longitudinal
velocity field $\delta_\tau v_{||}$ in a high Reynolds number turbulence experiment (see text).
Scales $\tau$ extend from the Kolmogorov scale to the integral scale.}
\end{figure}

Since our goal is to study the stationary random components of the
velocity field, we have preprocessed all time series in order to remove
the seasonal components. In fact the diurnal oscillations $S_{x,y}$
are not really periodic but vary during the year according to
the sun position. We have used a local harmonic parametrization
of these components and computed the parameters by minimizing
an exponential moving average of a quadratic error
(see \cite{Bovas,BaileInPrep} for more details).
In order to study the fluctuations in these
deseasonalized series and an eventual intermittency,
we can compute, as in turbulence literature, its increments. Notice that
one can alternaltively study velocity component wavelet coefficients or, if one wants to account for the low-frequency behavior,
the error in the one step forward prediction of a Langevin like modeling of $v_{x}(t)$ and $v_{y}(t)$ \cite{BaileInPrep}:
\begin{equation}
\label{llangevin}
   v_{x,y}(t+1) = S_{x,y}(t+1) + (1-\gamma) \left(v_{x,y}(t)-S_{x,y}(t)\right) + w_{x,y}(t)
\end{equation}
where the friction coefficient $\gamma$ is estimated to be
$\gamma \simeq 1$ day$^{-1}$, $S_{x,y}(t)$ stands for the seasonal
additive component of the wind velocity and $w_{x,y}(t)$ is the error (noise) term.
Whatever the precise definition used to compute the local fluctuations, the results presented in the next section remain unchanged.

\subsection{Evidences for a mesoscale cascade}

As recalled in the introduction, unlike inertial subrange turbulence,
few papers have been devoted to the study of scaling properties of atmospheric fields from
moderate to large scales. In Ref. \cite{lss}, the authors find that
cloud satellite data display multiscaling over the range $1-5.10^3$ km and suggest the existence of a (anisotropic)
cascade process from planetary scales down
to the microscales. As far as scaling properties of surface layer wind speed
are concerned, some studies focus on (multi-) scaling properties.
In Ref. \cite{kava04}, the authors use Detrented Fluctuation Analysis
on various hourly averaged wind series and provide evidences of a cross-over
between two scaling regimes separating mesoscale range from very large
scale range. The scale of the cross-over was found to be $T \simeq 5$ days
and the author conjectured the possibility of the existence of
multiscaling for scales below this scale. In that case the scale $T$ could
be identified to some integral (injection) scale.
In the Refs. \cite{lmsa,lma}, Lauren {\em et al.} perform an analysis and
a modeling of atmospheric wind in both
mesoscale and microscale (turbulent) regimes. They show that the
concepts of multiscaling and cascades are also pertinent for characterizing
and simulating low-wavenumbers properties of surface layer winds.

Though the above mentionned studies provide hints for the existence of intermittency
in wind fluctuations at large scales, as explained in section \ref{s_ma}, because of diurnal oscillations,
discreteness of the data, limited size of the scaling range and for purely statistical considerations,
the reported estimations of (multi-) scaling
properties of mesoscale wind data are far from being reliable.
A more direct and convincing illustration of the intermittent nature
of mesoscale wind variations is provided in Fig. \ref{figpdf}(a) where we have
plotted the normalized pdf of the increments of (deseasonalized) wind components $v_x$ of
the Eindhoven site in logaritmic scale. One sees that increment distributions go from `stretched exponential" like functions
(with large tails) to Gaussian like behavior (parabola) when going from 1 hour to few days time scales.
This behavior is very similar to the features observed in fully
developed turbulence laboratory experiments. In Fig. \ref{figpdf}(b), for comparison purpose, we have plotted
in the same logarithmic scale, the standardized pdf of longitudinal velocity increments
computed from experimental data obtained
by B. Chabaud and B. Castaing in a low temperature gaseous Helium
jet experiment \cite{chan00} (the Taylor scale based Reynolds
number is $R_\lambda = 929$). As the scale $\tau$ varies from the dissipation to the integral scale, one
observes the same deformation of the pdf from large tailed shape to Gaussian-like shape.
As explained is section \ref{s_ma1} (Eq. \eqref{castaing}), such a variation of the pdf behavior
across scales is usually associated with the existence of an intermittent cascade.
Let us notice that the smallest time scales we considered, correspond to scales greater than or equal to the
``injection'' largest scale of atmospheric boundary layer turbulence. Consequently, according to our observations,
velocity increment statistics at large (atmospheric surface layer) turbulent scales are characterized by a strong kurtosis.
This contrasts with the situation in laboratory experiments where the distribution are nearly gaussian at large scales (as illustrated by the bottom graph in Fig. \ref{figpdf}(b)).
Similar observations have been performed in \cite{bobape06,KhoMoTsi07}.
In \cite{bobape06}, the authors interpret the intermittency of large scale atmospheric turbulence by the fluctuations
of the turbulence intensity at this scale. The velocity mean is no longer constant but stochastic. Similar observations on the vorticity field
have been made in \cite{KhoMoTsi07}. In the following, we suggest that
these fluctuations of turbulence intensity can be interpreted as the result of a cascading process starting at a larger time scale.

As advocated in section \ref{s_ie}, the best way to reveal the presence of an underlying random cascade
and to estimate the intermittency coefficient is to study the magnitude correlations functions associated
with velocity small scale variations. If one writes equation \eqref{ss2} for the small scale increments of respectively
$v_x$ and $v_y$ (where the seasonal components have been removed) one can define two magnitude processes and
two processes $\Omega_x$ and $\Omega_y$:
\begin{eqnarray*}
\omega_{x,\tau}(t)  & \equiv & \ln(|\delta_\tau v_x (t)|) =  \Omega_{x,\tau}(t) + \ln(|\epsilon_x(t)|) \\
\omega_{y,\tau}(t)  &  \equiv & \ln(|\delta_\tau v_y (t)|) =  \Omega_{y,\tau}(t) + \ln(|\epsilon_y(t)|)
\end{eqnarray*}
If one assumes that the noises $\epsilon_x$ and $\epsilon_y$ are independent gaussian white noises, then
thanks to the fact that $\mbox{Var}(\ln(|\epsilon|)) \simeq 1.23$, one can compute the correlation coefficient of
$\Omega_x$ and $\Omega_y$ from the correlation of $\omega_x$ and $\omega_y$. The obtained results for all
data series are summarized in table \ref{table2}. One clearly sees that the correlation coefficient of magnitude
processes is for all series close to 20 \% while the estimated coefficients for the $\Omega$ components are close to 1
\footnote{notice that estimated coefficients greater than 1 are probably due to our hypothesis concerning the normality of $\epsilon_{x,y}$; such an assumption can only be a rough approximation for some series because of finite size and granularity effects}. This result strongly suggests that the processes $\Omega_x$ and $\Omega_y$ are identical and therefore $\Omega$ is a scalar quantity. One thus has (up to season components):
\begin{eqnarray*}
 \delta_\tau v_x(t) & = & e^{\Omega_\tau(t)} \epsilon_x(t) \\
 \delta_\tau v_y(t) & = & e^{\Omega_\tau(t)} \epsilon_y(t) \; .
\end{eqnarray*}
The scalar $e^{\Omega_\tau(t)}$ is simply the (stochastic) amplitude of velocity fluctuation vector at scale $\tau$.
\begingroup
\squeezetable
\begin{table}[h]
\begin{center}
 \begin{tabular}{|c||c|c|}
  \hline
   Location & $\omega$ correlations & $\Omega$ correlations \\
  \hline
   Vignola (Ajaccio) & 0.19 & 1.16 \\
  \hline
   Ajaccio & 0.22 & 1.31 \\
  \hline
   Bastia &  0.20 & 1.10 \\
  \hline
   Calvi & 0.14 & 1.10 \\
  \hline
   Conca & 0.22 & 1.07 \\
  \hline
   Figari & 0.17 & 1.32 \\
  \hline
   Renno & 0.21 & 0.71 \\
  \hline
   Sampolo & 0.11 & 0.68 \\
  \hline
   Eindhoven & 0.22 & 1.01 \\
  \hline
   Ijmuiden & 0.22 & 0.99 \\
  \hline
   Schipol & 0.21 & 0.98 \\
  \hline
\end{tabular}
\end{center}
\caption{Correlation coefficients of magnitude components for the different sites}
\label{table2}
\end{table}
\endgroup

\begin{figure}
\begin{center}
\includegraphics[width=7cm]{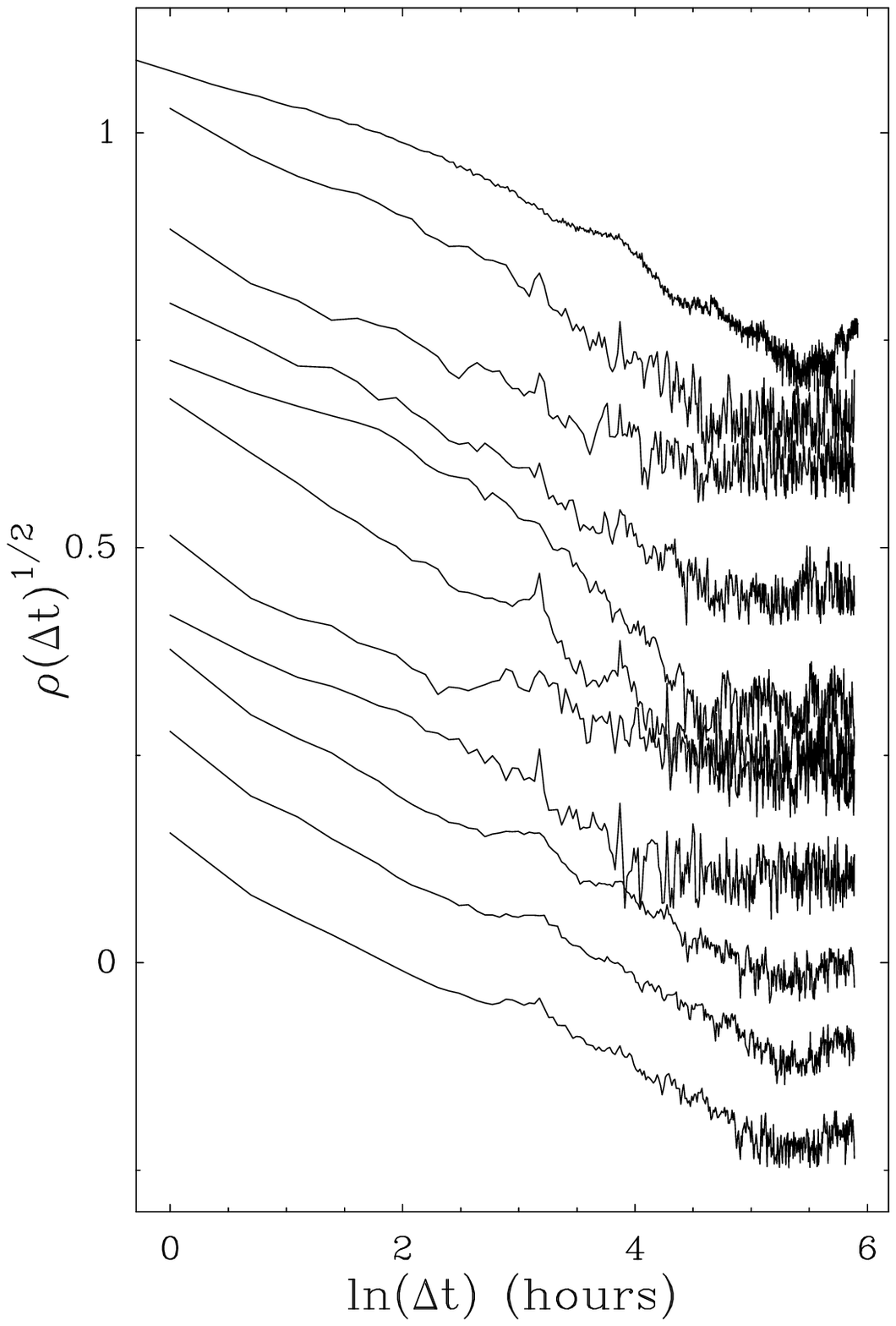}
\end{center}
\caption{Square root of magnitude covariance functions estimated for all wind series. The three bottom graphs correspond to Netherlands
hourly series while the top graph is correlation of magnitudes associated with the 'high frequency' Vignola data series. All the graphs
have been arbitrary shifted vertically for clarity purpose.
One can observe comparable values of the parameters $\beta^2$
and $T$ (see text).}
\label{fig_corr1}
\end{figure}

According to these findings, in the following we compute a surrogate of the scalar process $\Omega(t)$ as follows:
\begin{equation}
\label{somega}
  \Omega_\tau(t) = \frac{1}{2} \ln \left(\delta_\tau v_x (t)^2+\delta_\tau v_y (t)^2    \right)
\end{equation}
which is less noisy than individual magnitudes $\omega_x$ or $\omega_y$.
Let us notice, as already mentioned, that if one replaces in previous analysis, $\delta_\tau v_{x,y}$ by the
error term in a Langevin model of $v_{x,y}$ as in Eq. \eqref{llangevin} or by small scale wavelet coefficients,
all the results remain unchanged \cite{BaileInPrep}.
Since the results presented below do not depend on the chosen small scale $\tau$ (we focus on properties involving mainly lags
greater than $\tau$), we omit any reference to this scale
and denote $\Omega(t)$ the local ''magnitude'' estimated at small scale according to Eq. \eqref{somega}.

The process $\Omega(t)$ also possesses a seasonal component which means that seasonal effects
manifest not only through an additive (locally) periodic component but also through a diurnal modulation
of the velocity stochastic amplitude. This multiplicative seasonal component has been estimated using the same local least squared
method as for the additive components \cite{BaileInPrep}. In the following, we assume that all seasonal effects have been removed from
the estimated field $\Omega(t)$.

\begin{figure}
\begin{center}
\includegraphics[width=12cm]{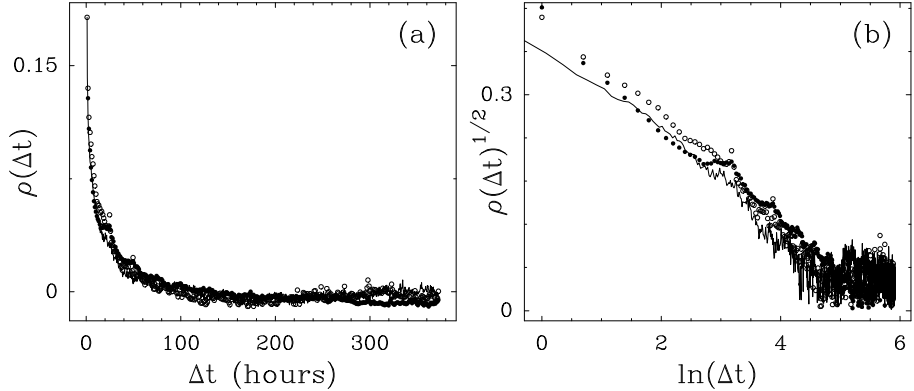}
\end{center}
\caption{Mean magnitude correlation functions associated with wind series in Corsica  ($\circ$) and Netherland ($\bullet$).
The solid line represents the magnitude covariance of the Vignola series (10 min rate). (a) The graphs are
in linear scales. (b) The square root of the covariances are represented as function of the logarithm of the time lag $\Delta t$.}
\label{fig_corr2}
\end{figure}

Along the line of the method described in section \ref{s_ie}, we have estimated the magnitude $\Omega(t)$
correlation functions $\rho(\Delta t) = \mbox{Cov}(\Omega(t),\Omega(t+\Delta t))$ for all the time series.
As illustrated in Fig. \ref{fig_corr1}, when one plots $\rho^{1/2}(\Delta t)$ as a function of $\ln(\Delta t)$ one observes, for all series, a decreasing linear function that becomes zero above some lag $T$. It is striking to observe that all slopes are close to each other and that the "correlation"
scale (integral scale) is roughly the same for all sites. In order to handle less noisy curves, we have plotted in Fig. \ref{fig_corr2}(b), within the
 same representation, the mean correlation function for all sites in Corsica, for all sites in Netherlands and for the high frequency series at Vignola (in Fig. \ref{fig_corr2}(a) the same graphs are displayed using a linear representation). Up the some remaining bump around the lag of one day due
 to the non perfect removing of the seasonal components, one observes a well defined linear dependence on a range $[\ln(\tau),\ln(T)]$.
 This means that for all wind series, the covariance of $\Omega(t)$ reads ($\Delta t > \tau$):
 \[
   \rho(\Delta t) = \beta^2 \ln^2(\frac{T}{\Delta t})
 \]
In section \ref{sq_mag}, we have explained how such a square logarithmic dependence of the single point covariance can be
the result of log-correlated Eulerian and Lagrangian fields. This feature has also
been observed on various laboratory turbulence data \cite{DelMuArn01,Cast02}.
These observations are therefore direct evidences that a random cascade mechanism can be involved in the energy transfer
at scales much greater than the usual turbulent microscale. Let us remark that the value of the integral scale we found
is $T \simeq 5$ days in Holland and Corsica and the value of the slope $\beta \simeq 0.07$. In order to deduce Eulerian (or
Lagrangian) value of the intermittency coefficients,
one would need to know the value of $\frac{T}{\eta}$
where $\eta$ is a small time scale cut-off
above which Lagrangian variations are no longer intermittent. For example if one sets $T \simeq 5$ days and one
chooses $\eta$ to be a typical time scale separating micro and meso scales, i.e., $\eta \simeq 10 \min$,
according to Eq. \eqref{l2sq} one gets an Eulerian intermittency coefficient $\lambda^2 \simeq 0.03$, i.e., very close to the value found
for fully developed turbulence (see e.g., \cite{ArnManMuzy98,Basu07,DelMuArn01}).

\section{Conclusion and prospects}
\label{s_cp}
The goal of this paper was to provide evidences that the surface layer wind fluctuation statistics in the mesoscale range are,
very much like microscale turbulent statistics, characterized by strongly intermittent properties.
We have first reviewed how the intermittency of continuous random cascades can be
advantageously studied using magnitude correlation analysis as compared to standard scaling or magnitude cumulant estimations.
Within this framework and using various wind velocity and direction time series in Corsica (France)
and Netherlands, we have shown that one can define a scalar magnitude field that displays
"universal" squared log correlation functions. Such peculiar shape of time correlation
functions at a fixed position
is shown to result from a continuous cascade (with log-correlated magnitude)
that governs the fluctuations in both
Eulerian and Lagrangian frames. The same behavior has been observed
in laboratory turbulence experiments.
The existence of some mesoscale "energy cascade" and other similarities
with the 3D isotropic turbulence properties, raises various fundamental questions. Unlike synoptic circulations,
mesoscale motions can be associated with a wide variety of phenomena covering a large
range of characteristic scales from thunderstorms, mountain waves to front dynamics. In that respect,
the ``universal'' features discussed in this paper
have to be confirmed using further experimental data. Let us note however
that a characteristic time scale of few days is usually associated with front dynamics \cite{dynmeteo}
and some comparable typical time scales have been already reported in the literature \cite{Gupta1,kava04}.
From a practical point of view our findings lead to a better characterization of the
statistical properties of wind fluctuations. The framework of intermittent statistics and
random cascade models can be applied to address problems related to wind resources assessing,
extreme events characterization or
to design a simple stochastic model in order to perform short term
wind predictions \cite{MuBaiPo09,BaileInPrep}.

\begin{acknowledgments}
We acknowledge ''Royal Netherlands Meteorological Institute'' \cite{KNMI} for the availability of the potential wind speed data in Netherlands
and B. Chabaud and B. Castaing for the permission to use their experimental turbulence data.
\end{acknowledgments}


\begin{thebibliography}{10}
\providecommand*{\bibinfo}[2]{#2}
\providecommand*{\eprint}[1]{#1}
\providecommand*{\url}[1]{#1}
\bibitem{Fri95}
\bibinfo{author}{U.~Frisch}, \bibinfo{title}{\emph{Turbulence}}
  (\bibinfo{publisher}{Cambridge Univ. Press}, Cambridge,
  \bibinfo{year}{1995}).
\bibitem{dynmeteo}
\bibinfo{author}{J.~Holton}, \bibinfo{title}{\emph{An introduction to Dynamic
  Meteorology}} (\bibinfo{publisher}{Elsevier Academic Press}, Burlington,
  \bibinfo{year}{2004}).
\bibitem{lmsa}
\bibinfo{author}{M.~K. Lauren}, \bibinfo{author}{M.~Menabde},
  \bibinfo{author}{A.~W. Seed}, and \bibinfo{author}{G.~Austin},
  \bibinfo{journal}{Boundary-Layer Meteorol.} \bibinfo{volume}{\textbf{90}},
  \bibinfo{pages}{21} (\bibinfo{date}{1999}).
\bibitem{lss}
\bibinfo{author}{S.~Lovejoy}, \bibinfo{author}{D.~Schwertzer}, and
  \bibinfo{author}{D.~Stanley}, \bibinfo{journal}{Phys. Rev. Lett.}
  \bibinfo{volume}{\textbf{86}}, \bibinfo{pages}{5200} (\bibinfo{date}{2001}).
\bibitem{Gupta1}
\bibinfo{author}{V.~Gupta} and \bibinfo{author}{E.~Waymire},
  \bibinfo{journal}{J. Geophys. Res.} \bibinfo{volume}{\textbf{95}},
  \bibinfo{pages}{1999} (\bibinfo{date}{1990}).
\bibitem{Gupta2}
\bibinfo{author}{V.~Gupta} and \bibinfo{author}{E.~Waymire},
  \bibinfo{journal}{J. Appl. Meteor.} \bibinfo{volume}{\textbf{32}},
  \bibinfo{pages}{251} (\bibinfo{date}{1993}).
\bibitem{kava04}
\bibinfo{author}{R.~G. Kavasseri} and \bibinfo{author}{R.~Nagarajan},
  \bibinfo{journal}{IEEE Trans. on Circuits and Systems. Fundam. Theory and
  Apps} \bibinfo{volume}{\textbf{51}}, \bibinfo{pages}{2262}
  (\bibinfo{date}{2004}).
\bibitem{bobape06}
\bibinfo{author}{F.~Bottcher}, \bibinfo{author}{S.~Barth}, and
  \bibinfo{author}{J.~Peinke}, \bibinfo{journal}{St. Env. Res. and Risk
  Assessment} \bibinfo{volume}{\textbf{21}}, \bibinfo{pages}{299}
  (\bibinfo{date}{2006}).
\bibitem{KhoMoTsi07}
\bibinfo{author}{M.~Kholomyansky}, \bibinfo{author}{L.~Moriconi}, and
  \bibinfo{author}{A.~Tsinober}, \bibinfo{journal}{Phys. Rev. E}
  \bibinfo{volume}{\textbf{76}}, \bibinfo{pages}{026307}
  (\bibinfo{date}{2007}).
\bibitem{ArnMuzSor98}
\bibinfo{author}{A.~Arneodo}, \bibinfo{author}{J.~F. Muzy}, and
  \bibinfo{author}{D.~Sornette}, \bibinfo{journal}{European Physical Journal B}
  \bibinfo{volume}{\textbf{2}}, \bibinfo{pages}{277} (\bibinfo{date}{1998}).
\bibitem{MuDeBa00}
\bibinfo{author}{J.-F. Muzy}, \bibinfo{author}{J.~Delour}, and
  \bibinfo{author}{E.~Bacry}, \bibinfo{journal}{Eur. J. Phys. B}
  \bibinfo{volume}{\textbf{17}}, \bibinfo{pages}{537} (\bibinfo{date}{2000}).
\bibitem{DelMuArn01}
\bibinfo{author}{J.~Delour}, \bibinfo{author}{J.~Muzy}, and
  \bibinfo{author}{A.~Arneodo}, \bibinfo{journal}{Eur. Phys. J. B}
  \bibinfo{volume}{\textbf{23}}, \bibinfo{pages}{243} (\bibinfo{date}{2001}).
\bibitem{BacKozMuz06}
\bibinfo{author}{E.~Bacry}, \bibinfo{author}{A.~Kozhemyak}, and
  \bibinfo{author}{J.~F. Muzy}, \bibinfo{journal}{Journal of Economic Dynamics
  and Control} \bibinfo{volume}{\textbf{32}}, \bibinfo{pages}{156}
  (\bibinfo{date}{2008}).
\bibitem{Koz06}
\bibinfo{author}{A.~Kozhemyak}, \bibinfo{title}{\emph{Mod\'elisation de
  s\'eries financi\`eres \`a l'aide de processus invariants d'\'echelle.
  Application \`a la pr\'ediction du risque}}, Ph.D. thesis, Ecole
  Polytechnique, Palaiseau, France (\bibinfo{date}{2006}).
\bibitem{DelourPHD}
\bibinfo{author}{J.~Delour}, \bibinfo{title}{\emph{Processus al\'eatoires
  auto-simimaires: applications en turbulence et en finance}}, Ph.D. thesis,
  Universit\'e de Bordeaux I, Pessac, France (\bibinfo{date}{2001}).
\bibitem{AnAnDa01}
\bibinfo{author}{F.~Anselmet}, \bibinfo{author}{R.~Antonia}, and
  \bibinfo{author}{L.~Danaila}, \bibinfo{journal}{Planetary and Space Science}
  \bibinfo{volume}{\textbf{49}}, \bibinfo{pages}{1177} (\bibinfo{date}{2001}).
\bibitem{LahAbrCha04}
\bibinfo{author}{B.~Lahermes}, \bibinfo{author}{P.~Abry}, and
  \bibinfo{author}{P.~Chainais}, \bibinfo{journal}{Int. J. of Wavelets,
  Multiresolution and Inf. Processing} \bibinfo{volume}{\textbf{2}}(4),
  \bibinfo{pages}{497} (\bibinfo{date}{2004}).
\bibitem{MuBaBaPo08}
\bibinfo{author}{J.~Muzy}, \bibinfo{author}{E.~Bacry},
  \bibinfo{author}{R.~Baile}, and \bibinfo{author}{P.~Poggi},
  \bibinfo{journal}{Europhys. Lett.} \bibinfo{volume}{\textbf{82}},
  \bibinfo{pages}{60007} (\bibinfo{date}{2008}).
\bibitem{Basu07}
\bibinfo{author}{S.~Basu}, \bibinfo{author}{E.~Foufoula-Georgiou},
  \bibinfo{author}{B.~Lashermes}, and \bibinfo{author}{A.~Arneodo},
  \bibinfo{journal}{Phys. Fluids} \bibinfo{volume}{\textbf{19}},
  \bibinfo{pages}{115102} (\bibinfo{date}{2007}).
\bibitem{ArnManMuzy98}
\bibinfo{author}{A.~Arneodo}, \bibinfo{author}{S.~Manneville}, and
  \bibinfo{author}{J.~Muzy}, \bibinfo{journal}{Eur. Phys. J. B}
  \bibinfo{volume}{\textbf{1}}, \bibinfo{pages}{129} (\bibinfo{date}{1998}).
\bibitem{CasGagHop90}
\bibinfo{author}{B.~Castaing}, \bibinfo{author}{Y.~Gagne}, and
  \bibinfo{author}{E.~Hopfinger}, \bibinfo{journal}{Physica D}
  \bibinfo{volume}{\textbf{46}}, \bibinfo{pages}{177} (\bibinfo{date}{1990}).
\bibitem{MuBa02}
\bibinfo{author}{J.-F. Muzy} and \bibinfo{author}{E.~Bacry},
  \bibinfo{journal}{Phys. Rev. E} \bibinfo{volume}{\textbf{66}},
  \bibinfo{pages}{056121} (\bibinfo{date}{2002}).
\bibitem{BaMu03}
\bibinfo{author}{E.~Bacry} and \bibinfo{author}{J.-F. Muzy},
  \bibinfo{journal}{Comm. in Math. Phys.} \bibinfo{volume}{\textbf{236}},
  \bibinfo{pages}{449} (\bibinfo{date}{2003}).
\bibitem{Man74a}
\bibinfo{author}{B.~B. Mandelbrot}, \bibinfo{journal}{Journal of Fluid
  Mechanics} \bibinfo{volume}{\textbf{62}}, \bibinfo{pages}{331}
  (\bibinfo{date}{1974}).
\bibitem{Man74b}
\bibinfo{author}{B.~B. Mandelbrot}, \bibinfo{journal}{C.R. Acad. Sci. Paris}
  \bibinfo{volume}{\textbf{278}}, \bibinfo{pages}{289} (\bibinfo{date}{1974}).
\bibitem{Man03}
\bibinfo{author}{B.~B. Mandelbrot}, \bibinfo{journal}{Jourrnal of Statistical
  Physics} \bibinfo{volume}{\textbf{110}}, \bibinfo{pages}{739}
  (\bibinfo{date}{2003}).
\bibitem{KaPe76}
\bibinfo{author}{J.~P. Kahane} and \bibinfo{author}{J.~Peyri\`ere},
  \bibinfo{journal}{Adv. in Mathematics} \bibinfo{volume}{\textbf{22}},
  \bibinfo{pages}{131} (\bibinfo{date}{1976}).
\bibitem{Gui87}
\bibinfo{author}{Y.~Guivarc'h}, \bibinfo{journal}{C.R. Acad. Sci. Paris}
  \bibinfo{volume}{\textbf{305}}, \bibinfo{pages}{139} (\bibinfo{date}{1987}).
\bibitem{Mol96}
\bibinfo{author}{G.~M. Molchan}, \bibinfo{journal}{Comm. in Math. Phys.}
  \bibinfo{volume}{\textbf{179}}, \bibinfo{pages}{681} (\bibinfo{date}{1996}).
\bibitem{Mol97}
\bibinfo{author}{G.~M. Molchan}, \bibinfo{journal}{Phys. Fluids}
  \bibinfo{volume}{\textbf{9}}, \bibinfo{pages}{2387} (\bibinfo{date}{1997}).
\bibitem{Liu02}
\bibinfo{author}{Q.~Liu}, \bibinfo{journal}{Asian Journal of Mathematics}
  \bibinfo{volume}{\textbf{6}}, \bibinfo{pages}{145} (\bibinfo{date}{2002}).
\bibitem{SchMar01}
\bibinfo{author}{F.~Schmitt} and \bibinfo{author}{D.~Marsan},
  \bibinfo{journal}{European Physical Journal B} \bibinfo{volume}{\textbf{20}},
  \bibinfo{pages}{3} (\bibinfo{date}{2001}).
\bibitem{BaMan02}
\bibinfo{author}{J.~Barral} and \bibinfo{author}{B.~B. Mandelbrot},
  \bibinfo{journal}{Prob. Theory and Relat. Fields}
  \bibinfo{volume}{\textbf{124}}, \bibinfo{pages}{409} (\bibinfo{date}{2002}).
\bibitem{ChaRieAbr05}
\bibinfo{author}{P.~Chainais}, \bibinfo{author}{R.~Riedi}, and
  \bibinfo{author}{P.~Abry}, \bibinfo{journal}{IEEE transactions on Information
  Theory} \bibinfo{volume}{\textbf{51}}, \bibinfo{pages}{1063}
  (\bibinfo{date}{2005}).
\bibitem{ArnBaMaMu98}
\bibinfo{author}{A.~Arneodo}, \bibinfo{author}{E.~Bacry},
  \bibinfo{author}{S.~Manneville}, and \bibinfo{author}{J.~Muzy},
  \bibinfo{journal}{Phys. Rev. Lett.} \bibinfo{volume}{\textbf{80}},
  \bibinfo{pages}{708} (\bibinfo{date}{1998}).
\bibitem{MuzDelBac00}
\bibinfo{author}{J.~F. Muzy}, \bibinfo{author}{J.~Delour}, and
  \bibinfo{author}{E.~Bacry}, \bibinfo{journal}{Eur. J. Phys. B}
  \bibinfo{volume}{\textbf{17}}, \bibinfo{pages}{537} (\bibinfo{date}{2000}).
\bibitem{BacKozMuz09}
\bibinfo{author}{E.~Bacry}, \bibinfo{author}{A.~Kozhemyak}, and
  \bibinfo{author}{J.~F. Muzy}, \bibinfo{title}{\emph{Log-normal continuous
  cascades: aggregation properties and estimation}} (\bibinfo{date}{2009}),
  quantitative Finance (in press).
\bibitem{audit}
\bibinfo{author}{B.~Audit}, \bibinfo{author}{E.~Bacry},
  \bibinfo{author}{J.~Muzy}, and \bibinfo{author}{A.~Arneodo},
  \bibinfo{journal}{IEEE Trans. Info. Theory} \bibinfo{volume}{\textbf{48}},
  \bibinfo{pages}{2938} (\bibinfo{date}{2002}).
\bibitem{Che05}
\bibinfo{author}{L.~Chevillard}, \bibinfo{author}{S.~Roux},
  \bibinfo{author}{E.~Leveque}, \bibinfo{author}{N.~Mordant},
  \bibinfo{author}{J.~Pinton}, and \bibinfo{author}{A.~Arneodo},
  \bibinfo{journal}{Phys. Rev. Lett.} \bibinfo{volume}{\textbf{95}},
  \bibinfo{pages}{064501.1} (\bibinfo{date}{2005}).
\bibitem{Cast02}
\bibinfo{author}{B.~Castaing}, \bibinfo{journal}{Eur. Phys. J. B}
  \bibinfo{volume}{\textbf{29}}, \bibinfo{pages}{357} (\bibinfo{date}{2002}).
\bibitem{Cast06}
\bibinfo{author}{B.~Castaing}, \bibinfo{journal}{Phys. Rev. E}
  \bibinfo{volume}{\textbf{73}}, \bibinfo{pages}{068301}
  (\bibinfo{date}{2006}).
\bibitem{KNMI}
See www.knmi.nl/samenw/hydra/index.html.
\bibitem{vdh}
\bibinfo{author}{I.~V. der Hoven}, \bibinfo{journal}{J. of Meteorol.}
  \bibinfo{volume}{\textbf{14}}, \bibinfo{pages}{160} (\bibinfo{date}{1957}).
\bibitem{ot}
\bibinfo{author}{A.~H. Oort} and \bibinfo{author}{A.~Taylor},
  \bibinfo{journal}{Month. Weather Rev} \bibinfo{volume}{\textbf{97}},
  \bibinfo{pages}{623} (\bibinfo{date}{1969}).
\bibitem{NasGage83}
\bibinfo{author}{G.~Nastrom}, \bibinfo{author}{K.~Gage}, and
  \bibinfo{author}{W.~Jasperson}, \bibinfo{journal}{Nature}
  \bibinfo{volume}{\textbf{310}} (\bibinfo{date}{1983}).
\bibitem{Lilly83}
\bibinfo{author}{D.~Lilly}, \bibinfo{journal}{J. Atmos. Sci.}
  \bibinfo{volume}{\textbf{40}} (\bibinfo{date}{1983}).
\bibitem{kaimal}
\bibinfo{author}{J.~Kaimal}, \bibinfo{journal}{J. Atmos. Sci.}
  \bibinfo{volume}{\textbf{35}}, \bibinfo{pages}{18} (\bibinfo{date}{1978}).
\bibitem{Taqqu}
\bibinfo{author}{M.~Taqqu} and \bibinfo{author}{G.~Samorodnisky},
  \bibinfo{title}{\emph{Stable Non-Gaussian Random Processes}}
  (\bibinfo{publisher}{Chapman \& Hall}, New-York, \bibinfo{year}{1994}).
\bibitem{Bovas}
\bibinfo{author}{B.~Abraham} and \bibinfo{author}{J.~Ledolter},
  \bibinfo{title}{\emph{Statistical methods for forectasting}}
  (\bibinfo{publisher}{J. Wiley \& Sons}, New-York, \bibinfo{year}{1983}).
\bibitem{BaileInPrep}
\bibinfo{author}{R.~Baile}, \bibinfo{author}{J.~Muzy}, and
  \bibinfo{author}{P.~Poggi}, \bibinfo{title}{\emph{Short term wind speed
  forecasting using continuous cascade model}} (\bibinfo{date}{2009}),
  preprint.
\bibitem{lma}
\bibinfo{author}{M.~K. Lauren}, \bibinfo{author}{M.~Menabde}, and
  \bibinfo{author}{L.~Austin}, \bibinfo{journal}{Boundary-Layer Meteorol.}
  \bibinfo{volume}{\textbf{100}}, \bibinfo{pages}{263} (\bibinfo{date}{2001}).
\bibitem{chan00}
\bibinfo{author}{O.~Chanal}, \bibinfo{author}{B.~Chabaud},
  \bibinfo{author}{B.~Castaing}, and \bibinfo{author}{B.~Hebral},
  \bibinfo{journal}{Eur. Phys. J. B} \bibinfo{volume}{\textbf{17}},
  \bibinfo{pages}{309} (\bibinfo{date}{2000}).
\bibitem{MuBaiPo09}
\bibinfo{author}{R.~Baile}, \bibinfo{author}{J.~Muzy}, and
  \bibinfo{author}{P.~Poggi}, \bibinfo{title}{\emph{Intermittency of surface
  layer wind speed fluctuations in the meso-scale range: application to short
  term prediction}} (\bibinfo{date}{2009}), to appear in Proceedings of World
  Renewable Energy Congress, Bangkok, Thailand, 2009.

\end{thebibliography}
\end{document}